\documentclass[aps,reprint,superscriptaddress]{revtex4-2}
\usepackage{amsmath}
\usepackage{amssymb}
\usepackage{graphicx}
\usepackage{orcidlink}
\usepackage{hyperref}
\usepackage{bm}
\usepackage{color}
\usepackage{pdfpages}

\hypersetup{colorlinks=true}

\makeatletter
\AtBeginDocument{\let\LS@rot\@undefined}
\makeatother

\begin{document}



\title{Detecting Linear Breit-Wheeler Signals with a Laser-Foil Setup}

\author{Huai-Hang Song
}\email{huaihangsong@sjtu.edu.cn}
\affiliation{Key Laboratory for Laser Plasmas (MOE) and School of Physics and Astronomy, Shanghai Jiao Tong University, Shanghai 200240, China}
\affiliation{Collaborative Innovation Center of IFSA, Shanghai Jiao Tong University, Shanghai 200240, China}

\author{Wei-Min Wang
}
\affiliation{Department of Physics and Beijing Key Laboratory of Opto-electronic Functional Materials and Micro–nano Devices, Renmin University of China, Beijing 100872, China}
\affiliation{Collaborative Innovation Center of IFSA, Shanghai Jiao Tong University, Shanghai 200240, China}

\author{Yu-Tong Li
}
\affiliation{Beijing National Laboratory for Condensed Matter Physics, Institute of Physics, Chinese Academy of Sciences, Beijing 100190, China}
\affiliation{School of Physical Sciences, University of Chinese Academy of Sciences, Beijing 100049, China}
\affiliation{Songshan Lake Materials Laboratory, Dongguan, Guangdong 523808, China}

\author{Yutong He}
\affiliation{Max-Planck-Institut für Kernphysik, Saupfercheckweg 1, 69117 Heidelberg, Germany}

\author{Matteo Tamburini}
\affiliation{Max-Planck-Institut für Kernphysik, Saupfercheckweg 1, 69117 Heidelberg, Germany}

\author{Christoph H. Keitel}
\affiliation{Max-Planck-Institut für Kernphysik, Saupfercheckweg 1, 69117 Heidelberg, Germany}

\author{Zheng-Ming Sheng
}
\email{zmsheng@sjtu.edu.cn}
\affiliation{Key Laboratory for Laser Plasmas (MOE) and School of Physics and Astronomy, Shanghai Jiao Tong University, Shanghai 200240, China}
\affiliation{Collaborative Innovation Center of IFSA, Shanghai Jiao Tong University, Shanghai 200240, China}
\affiliation{Tsung-Dao Lee Institute, Shanghai Jiao Tong University, Shanghai 201210, China}

\date{\today}

\begin{abstract}

As a fundamental QED process, linear Breit-Wheeler (LBW) pair production predicted 90 years ago has not yet been demonstrated in experiments with real photons. Here, we propose an experimentally advantageous scheme to detect the LBW signal by irradiating a foil target with a single 10 PW-level laser. Our integrated QED particle-in-cell simulations demonstrate that the LBW signal can be explicitly distinguished from the Bethe-Heitler (BH) signal by comparing positron energy spectra behind the target at varying target thicknesses. The LBW positrons are created at the front of the target and subsequently experience both laser vacuum acceleration and sheath field acceleration to gain high energies, while BH positrons, originating within the target bulk, are only subjected to sheath field acceleration. As a result, the invariance of the high-energy tail of positron spectra with respect to the target thickness serves as a distinct signature of the LBW process. Notably, this scheme remains viable even when the BH yield dominates over the LBW yield.

\end{abstract}


\maketitle


From the perspective of the number of absorbed photons, QED processes can be broadly categorized into linear QED processes and nonlinear QED processes \cite{berestetskii1982qed,jauch1976book,piazza2012rmp,gonoskov2022rmp,fedotov2023pr}. Nonlinear QED processes are the interaction of a high-energy lepton or photon with strong fields involving multiple low-energy photons \cite{ritus1985jslr,baier1998qed}, and many of them have been detected in experiments, like nonlinear Compton scattering (NCS) \cite{bula1996prl,mirzaie2024np} and nonlinear Breit-Wheeler (NBW) pair production \cite{burke1997prl}. Linear QED processes describe the collision of two single leptons or photons, whereas those associated with photon-photon collisions are significantly more challenging to detect \cite{ginzburg1981jetp,Lundstrom2006prl,sangal2021prd}.

As a basic linear QED process predicted in 1934 \cite{breit1934pr}, linear Breit-Wheeler (LBW) pair production refers to that an electron-positron ($e^\pm$) pair is produced via the collision of two high-energy photons. Although some experiments \cite{adam2021prl} have investigated it by quantizing photons from self-generated electromagnetic fields of ultrarelativistic charged nuclei \cite{weizsacker1934zfp}, its direct detection using real photons remains experimentally elusive. This challenge arises primarily from its small cross section ($\sim10^{-25}~\rm cm^2$) and high-energy threshold ($> 0.511$ MeV), which demand a brilliant source of high-energy photons—a capability that has been unavailable in laboratories for decades.

In recent years, the construction of 10 PW and even 100 PW lasers has surged globally \cite{tanaka2020mre,burdonov2021mre,yoon2021optica,li2022hplse}. These high-intensity lasers can be applied to effectively generate brilliant $\gamma$ photons through laser-plasma interactions via NCS \cite{nerush2011prl,ridgers2012prl,brady2012prl,ji2014prl,vranic2019pop,stark2016prl,gonoskov2017prx,wang2018pnas}. Accordingly, many approaches have been proposed to detect LBW signals with laser-plasma-driven $\gamma$ photons \cite{pike2014np,ribeyre2016pre,yu2019prl,wang2020prl,he2021cp,he2021njp,sugimoto2023prl}. Most of these methods require the collision of two collimated and brilliant $\gamma$-photon beams in vacuum, necessitating the use of two independent lasers, specialized microtargets, and high-precision alignment for the collision, which imposes considerable technical challenges. Recently, single-laser setups, offering greater experimental accessibility, have been numerically demonstrated to initiate LBW pair production in near-critical-density plasmas \cite{he2021njp,sugimoto2023prl}. In these setups, the emitted photons propagate both in forward and backward directions, forming a photon-photon collider. An accompanying issue is that, in plasma environments, the Bethe-Heitler (BH) process \cite{bethe1034prsa}---where photons decay into $e^\pm$ pairs in the Coulomb field of ions---becomes inevitable. Although simulations have suggested that LBW yields can exceed BH yields in near-critical-density plasmas, distinguishing between LBW and BH processes under these conditions remains a considerable difficulty. This stems from the similar post-production acceleration experienced by LBW and BH positrons, complicating their differentiation based on positron energy spectra.

\begin{figure}[t]
\centering
\includegraphics[width=\columnwidth]{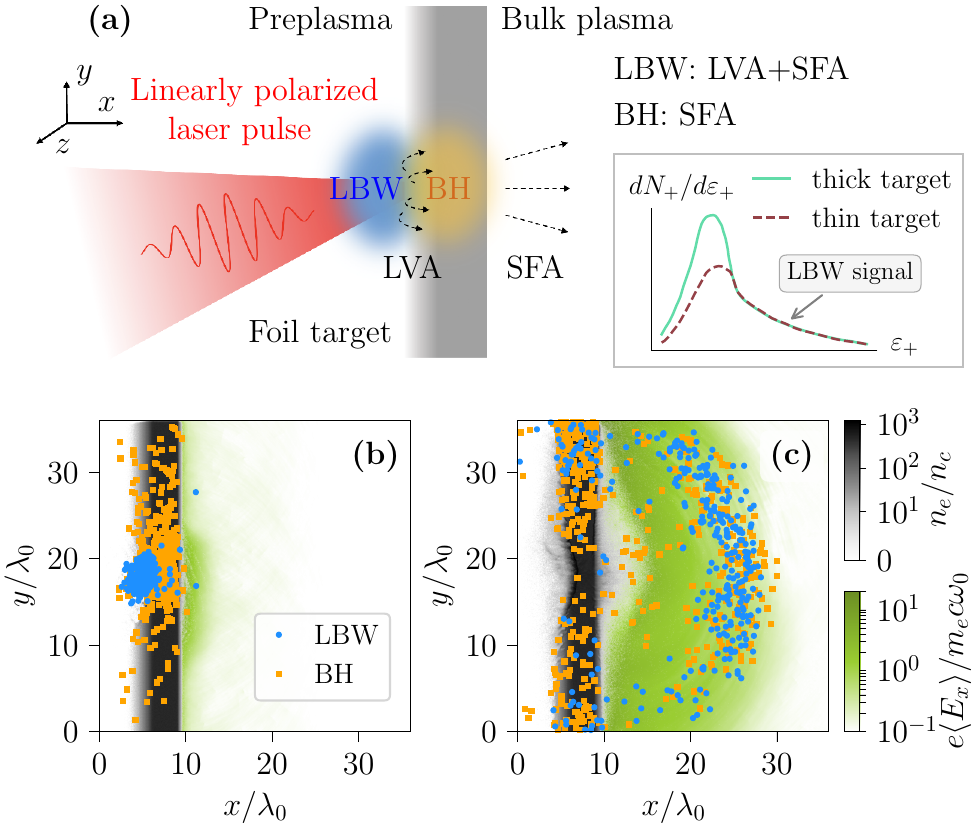}
\caption{\label{fig1} (a) Schematic of the proposed scheme. A 10 PW-class linearly polarized laser pulse irradiates a foil target, with some preplasmas positioned at the front of the target. The blue and orange regions indicate the respective creation sites of LBW and BH positrons. The typical energy spectra of the total (LBW and BH) positrons detected behind the target for thin and thick targets are illustrated in the inset, where the invariance of high-energy positrons serves as a LBW signature. (b, c) Simulation results showing the one-laser-cycle-averaged sheath electric field $\langle E_x \rangle$ behind the target and the electron density $n_e$ at $t = 20T_0$ (b) and $t = 40T_0$ (c), where the positions of some LBW and BH positrons are also illustrated at their creation (b) and $t = 40T_0$ (c), respectively.}
\end{figure}

In this Letter, we propose a method to detect and distinguish LBW signals from other pair production mechanisms using a conventional laser-foil setup, in which a single 10 PW-level linearly polarized laser irradiates a foil target, as depicted in Fig.~\ref{fig1}(a). Some low-density preplasmas, naturally formed by laser prepulses in real experiments, are introduced at the front of the target. The laser interacts strongly with these preplasmas, accelerating electrons to hundreds of MeV and generating high-energy $\gamma$ photons through NCS both in forward and backward directions. These near-isotropically emitted $\gamma$ photons can facilitate photon-photon collisions near the front surface of the target, resulting in substantial LBW pair production \cite{song2024pre}. Meanwhile, forward-directed photons can also decay into pairs via the BH process as they propagate through the bulk plasma and collide with ions. The distinct generation locations of LBW and BH positrons give rise to differences in their final energy spectra and dependence on the target thickness. The LBW positrons experience both laser vacuum acceleration (LVA) at the front of the target and sheath field acceleration (SFA) behind it, while the BH positrons undergo SFA only. Consequently, increasing the target thickness leads to a rise in intermediate-energy positrons, which serves as the BH signature, whereas the number of high-energy positrons, indicative of the LBW process, remains unchanged [see the insert of Fig.~\ref{fig1}(a)].

We have validated our scheme through a series of two-dimensional (2D)
particle-in-cell (PIC) simulations using our {\scshape yunic} code \cite{song2021arxiv}. To our knowledge, this is the first time to self-consistently simulate LBW, BH and NBW pair production with weighted macroparticles that are commonly used in modern PIC codes \cite{higginson2019jcp}. In our QED module, two nonlinear processes---NCS and NBW pair production---are modeled with a standard Monte-Carlo method under the locally constant field approximation \cite{elkina2011prab,ridgers2014jcp,gonoskov2015pre,song2021arxiv,montefiori2023cpc}. Both LBW and BH processes based on pairwise collisions fully account 
for angle-resolved $e^\pm$ emission, with implementation details and 
benchmarks provided in \cite{song2024pre} and the supplemental material \cite{material}. A 2D PIC simulation with an independent implementation of the LBW 
process in the {\scshape epoch}
code \cite{arber2015ppcf} successfully reproduced our LBW results \cite{material}. Moreover, a fully three-dimensional PIC simulation with the {\scshape yunic} code has also confirmed our 2D results \cite{material}. The bremsstrahlung is neglected here, as we found through additional simulations conducted with the {\scshape epoch} code that its intensity is 2 to 3 orders of magnitude lower than that of NCS under our representative parameters. Other pair production mechanisms related to the Coulomb field also do not affect the high-energy LBW signal, as they are generated within the target and contribute only to low- and intermediate-energy positrons.

In a representative simulation, a fully ionized carbon foil target ($Z=6$) is placed at $x=7.5~\mu$m with a thickness of $3~\mu$m. The bulk plasma has an electron density of $400n_c$, where $n_c=m_e\omega_0^2/4\pi e^2$ is the critical plasma density, $\omega_0$ is the laser angular frequency, $m_e$ is the electron mass, and $e$ is the elementary charge. The preplasma in front of the target features an exponential density profile with a scale length of $L_0^{\rm front}=0.5~\mu$m. A linearly polarized laser pulse, polarized in the $x$-$y$ plane, is incident from the left boundary ($x=0$) at an angle of $20^\circ$ relative to the target's normal direction. The laser has a central wavelength of $\lambda_0=1~\mu$m, a normalized amplitude of $a_0=eE_0/m_ec\omega_0=140$ (corresponding to a laser intensity of $2.7\times10^{22}~\rm W/cm^2$), a FWHM duration of $6T_0$, and a waist radius of $3\lambda_0$, where $T_0=\lambda_0/c\approx3.3$ fs and $c$ is the light speed in vacuum. The simulation domain measures $L_x\times L_y=36\lambda_0\times36\lambda_0$ and is resolved with $1440\times1440$ cells. Each cell contains 64 electrons and 64 carbon ions. To reduce Monte-Carlo noise, we performed the simulations three times for each case using different random seeds and averaged the results to ensure statistical accuracy. The positron energy spectrum and yield discussed throughout this Letter consider only positrons detected behind the target. The real positron number is obtained by assuming the missed dimension length along the $z$ direction to be $1~\mu$m. At this laser intensity, no NBW positrons are observed in simulations; therefore, our analysis first focuses on LBW and BH positrons.

Our simulation results in Fig.~\ref{fig1}(b) confirm that LBW positrons are predominantly created at the front of the target, while BH positrons are primarily generated in the bulk plasma. This distinction arises from the different particle 
species involved in two processes. In the LBW process, positrons are produced by photon-photon collisions, i.e., $\gamma+\gamma\rightarrow e^-+e^+$, which is most efficient when the photons collide head-on with each other, and less favorable when they tend to copropagate. As a result, LBW positrons are mainly created at the front of the target, where photons are emitted in both forward and backward directions with a wide angular distribution \cite{song2024pre}. When forward-propagating photons travel through the bulk plasma, LBW pair production is suppressed due to the nearly collinear photon propagation. In contrast, BH pairs are predominantly produced in the bulk plasma through collisions between forward photons and nearly rest carbon ions, i.e., $\gamma+Z\rightarrow e^-+e^+$, thus away from the influence of the laser field.

\begin{figure}[t]
\centering
\includegraphics[width=\columnwidth]{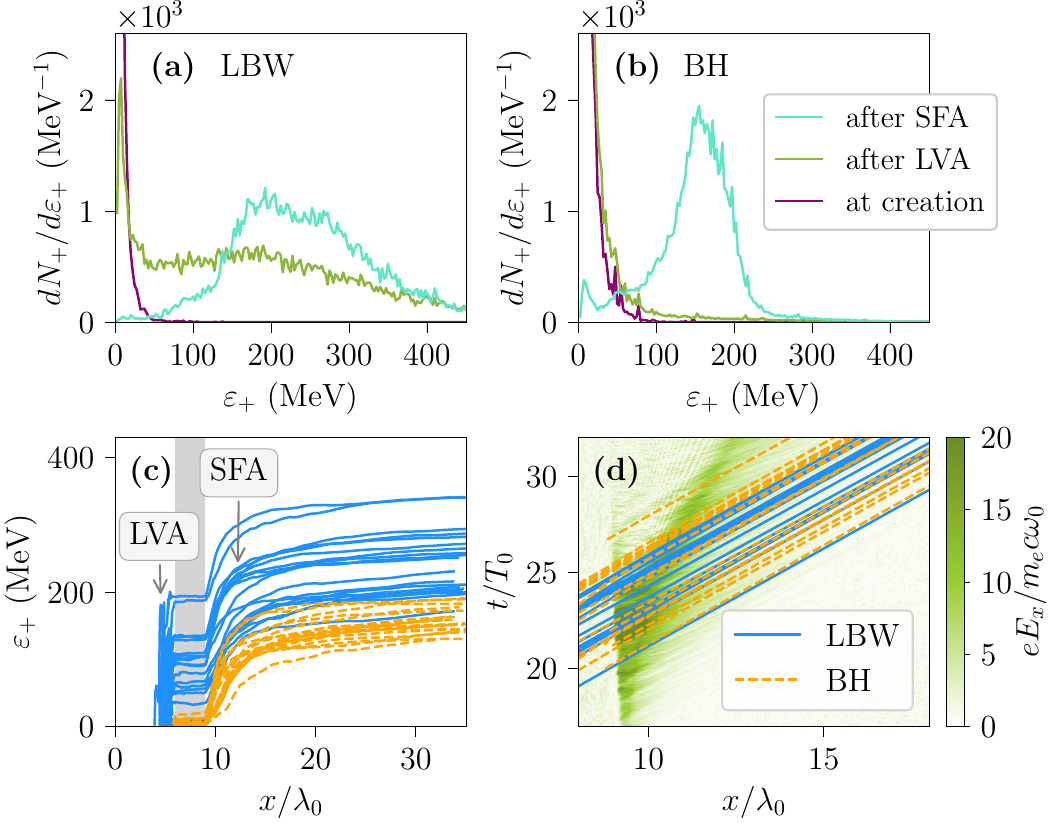}
\caption{\label{fig2} (a, b) Energy spectra of LBW positrons (a) and BH positrons (b) recorded at three stages: at creation, after LVA (but before SFA), and after SFA (also after LVA). (c) The energy $\varepsilon_+$ versus the position $x$ of some tracked LBW and BH positrons, with the gray-shaded region indicating the target location. (d) The spatialtemporal evolution of the sheath electric field $E_x$ and the tracked LBW and BH positrons.}
\end{figure}

In later times, a large number of both LBW and BH positrons appear behind the target, as shown in Fig.~\ref{fig1}(c). After being generated, LBW positrons are pushed forward into the bulk plasma by the strong laser field via LVA and subsequently escape from the back surface of the target. On the other hand, BH positrons are unaffected by the laser field as they are primarily produced in the bulk plasma; they move forward because their initial momenta, inherited from the photons, is directed forward. Although some BH positrons can pass through the target, a portion of them are captured due to their low energies. It is hard to distinguish LBW and BH positrons based on their angular distribution, as both of them predominantly propagate in the forward direction. Behind the target, a strong sheath field composed of quasistatic electric fields along the $+x$ direction [Figs.~\ref{fig1}(b) and \ref{fig1}(c)] is excited by the escaping of fast electrons from the target plasma. We will demonstrate that positron acceleration by the laser field and sheath field play the crucial role in the LBW detection.

The LBW and BH positrons undergo distinct acceleration processes, as evident by their energy spectra shown in Figs.~\ref{fig2}(a) and \ref{fig2}(b). Initially, both LBW and BH positrons have low energies, typically below 20 MeV, seen from the at-creation spectra. This is because the photon decay probability in both LBW and BH pair production is only weakly dependent on the photon energy, in contrast to the NBW process that will be discussed later. Thus, most of positrons are generated by relatively low-energy photons, as they are predominantly radiated in NCS. \emph{The LBW positrons generated at the front of the target undergo strong LVA, whereas BH positrons, created in the bulk plasma, do not}. Following LVA, LBW positrons are accelerated to high energies, with a broad spectrum and a maximum energy exceeding 450 MeV, as indicated by the after-LVA spectrum in Fig.~\ref{fig2}(a). The LVA does not result in a clear peak feature in the LBW positron 
spectrum. Both LBW and BH positrons then experience SFA behind the target, resulting in an energy peak of 100--200 MeV, as shown by the after-SFA spectra. The acceleration dynamics is further illustrated by the position-energy evolution of some tracked positrons in Fig.~\ref{fig2}(c). It clearly shows that the LBW positrons experience two acceleration stages: LVA at the target front ($x \lesssim 6~\mu$m) and SFA in the target back ($x > 9~\mu$m), while the tracked BH positrons only undergo SFA. The spatialtemporal evolution of the longitudinal electric field and the trajectories of tracked positrons are illustrated in Fig.~\ref{fig2}(d). The generation of positrons and the excitation of quasistatic electric fields occur synchronously. The positrons appear as a bunch passing through the sheath field, gaining energy to achieve a quasimonoenergetic peak. This contrasts sharply with the exponentially dropping energy spectrum of ions via thermal plasma-vacuum expansion  (target normal sheath acceleration of ions \cite{macchi2013rmp}). An energy peak at several MeV for BH positrons has been observed in experiments using $10^{19}$--$10^{20}~\rm W/cm^2$, ps lasers \cite{chen2009prl,yan2017ppcf}. Our findings demonstrate that quasimonoenergetic positrons accelerated by the sheath field remain feasible with 10 PW-level, fs lasers, achieving peak energies exceeding 100 MeV.

\begin{figure}[t]
\centering
\includegraphics[width=\columnwidth]{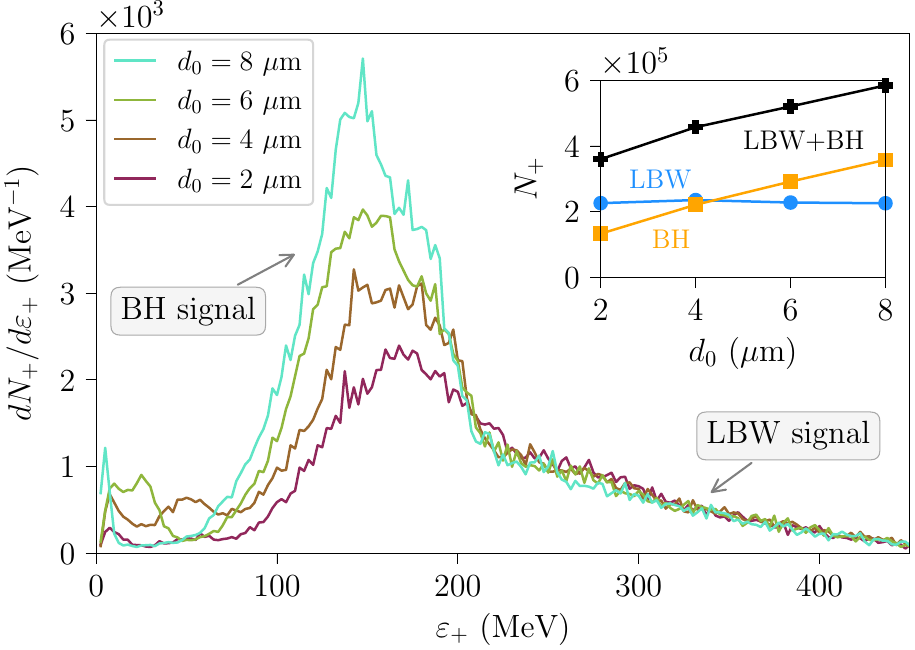}
\caption{\label{fig3} Energy spectra of total (LBW and BH) positrons measured behind the target at the end of the interaction for four different target thicknesses. The inset is the yield of LBW, BH and total positrons as a function of the target thickness. All other parameters are the same as those in Fig.~\ref{fig2}.}
\end{figure}

The difference in acceleration processes between LBW and BH positrons is crucial for distinguishing them, particularly through the comparison of positron spectra detected behind the target at varying target thicknesses, as illustrated in Fig.~\ref{fig3}. When the target thickness increases from 2 $\mu$m to 8 $\mu$m, high-energy positrons ($> 250$ MeV) remain unaffected, whereas intermediate-energy positrons (100--200 MeV) exhibit a strong sensitivity to the target thickness. As previously discussed, LBW positrons undergo both LVA and SFA, achieving higher energies than BH positrons. \emph{These high-energy positrons, whose yield remains unaffected with the target thickness, serve as a definitive signature of LBW pair production}. The influence of target thickness on the low-energy positrons ($<90$ MeV) appears somewhat irregular, as they are highly sensitive to the sheath field, which is in turn affected by the target thickness. The inset of Fig.~\ref{fig3} illustrates the yield of LBW positrons, BH positrons, and the total (LBW and BH) positrons as functions of the target thickness. The number of LBW positrons remains nearly constant across the scanned target thicknesses, while the number of BH positrons increases approximately linearly, as BH pair production is driven by photon-ion collisions. From an experimental perspective, isolating LBW signals based solely on the total positron yield is challenging, as the total yield always increases with the target thickness. For ultrathin targets, the LBW positron yield is also influenced by the target thickness, as the intense laser can drive the target plasma into relativistic transparency. Importantly, our approach does not require LBW positrons to dominate. Even for targets much thicker than $8~\mu$m, high-energy LBW positrons remain distinguishable, enabling effective identification of LBW signals.

\begin{figure}[t]
\centering
\includegraphics[width=\columnwidth]{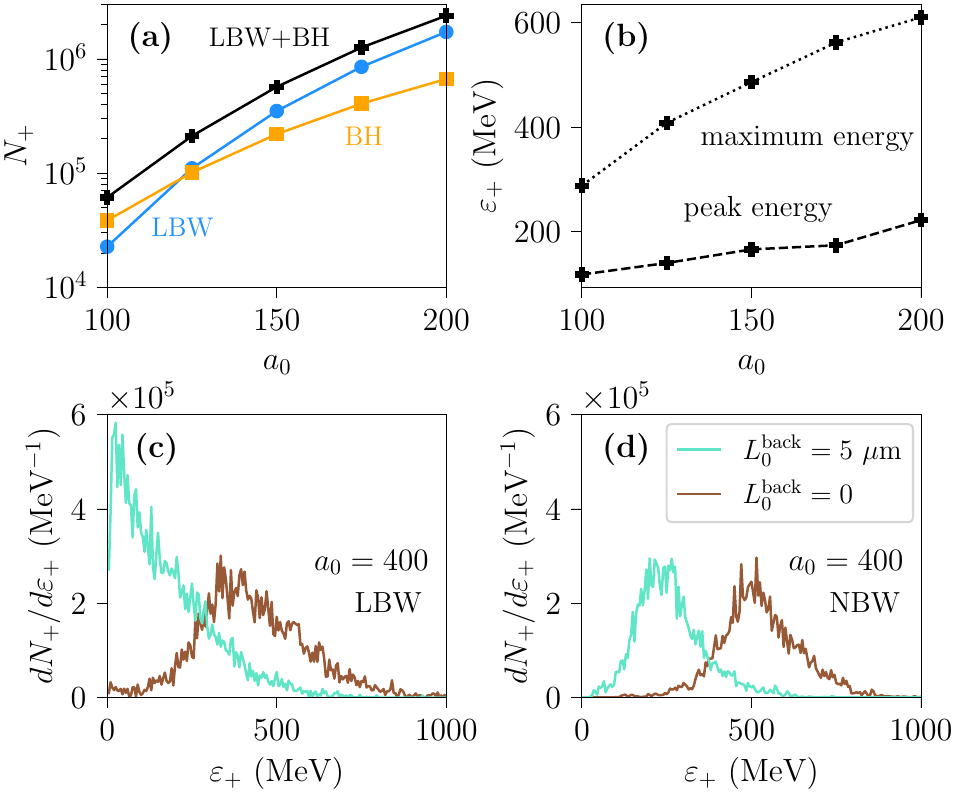}
\caption{\label{fig4} (a, b) Influence of the laser intensity on the yield of  LBW, BH, and total (LBW and BH) positrons (a), and the peak and maximum energies of the total positrons (b) at $a_0=100$--200. (c, d) Energy spectra of LBW positrons (c) and NBW positrons (d) without ($L_0^{\rm back}=0$) and with ($L_0^{\rm back}=5~\mu$m) back-target preplasmas at a high laser intensity $a_0=400$ and a thick target of the $6~\mu$m thickness. All other parameters are the same as those in Fig.~\ref{fig2}.}
\end{figure}

The influence of laser intensities on the LBW and BH positrons is summarized in Figs.~\ref{fig4}(a) and \ref{fig4}(b). At laser intensities of $a_0>120$, the LBW process surpasses the BH process [Fig.~\ref{fig4}(a)]. This can be attributed to the dependence of the positron yield on particle densities: $dN_+^{\rm LBW}/dt\propto n_\gamma^2$ for LBW positrons, and $dN_+^{\rm BH}/dt \propto n_\gamma n_i$ for BH positrons, where $n_\gamma$ and $n_i$ are the photon density and ion density. As the laser intensity increases, $n_\gamma$ rises significantly, finally leading to a higher LBW yield compared to the BH yield. Figure~\ref{fig4}(b) illustrates the peak and maximum energies of the total positrons as functions of laser intensities. By such a typical laser-foil setup of the normalized laser amplitude $a_0=100$--200, quasimonoenergetic positrons with peak energies of approximately 100--200 MeV and maximum energies ranging from 200--600 MeV can be generated. The maximum positron energy is about three times higher than the peak energy, indicating that the high-energy LBW signal can be maintained across a broad range of laser intensities.

Another important pair production mechanism---the NBW process---becomes increasingly significant at high laser intensities of $a_0>300$. The dominant relationship between LBW and NBW processes has been thoroughly discussed in the previous work \cite{song2024pre}. Since NBW positrons are also generated at the front of the target, they are likewise unaffected by the target thickness. Here, we clarify that LBW signals can be distinguished from NBW signals through relatively low-energy positrons. The LBW positrons are primarily generated by relatively low-energy photons, resulting in their low initial energies. In contrast, NBW positrons require sufficiently high photon energies, as low-energy photons are exponentially suppressed in their NBW pair production. This results in a Gaussian-shape NBW spectrum with negligible low-energy positrons. To minimize the influence of subsequent SFA, some additional preplasmas with a scale length of $L_0^{\rm back}=5~\mu$m are also placed at the back of the target to suppress the sheath field, which is a widely used experimental technique \cite{mackinnon2001prl}. Since no significant NBW pair production is observed at $a_0<200$, we demonstrate it at a high laser intensity of $a_0=400$ and a corresponding thick target thickness of $d_0=6~\mu$m. As shown in Figs.~\ref{fig4}(c) and \ref{fig4}(d), LBW positrons are dominated by low-energy positrons, while NBW positrons contain only a small fraction of low-energy positrons when $L_0^{\rm back}=5~\mu$m. In comparison, if there are no back-target preplasmas, i.e., $L_0^{\rm back}=0$, LBW positrons will be accelerated by the sheath field to high energies with a peak at about 350 MeV, displaying a similar Gaussian-shape spectrum to that of NBW positrons.

In conclusion, we have proposed a straightforward scheme to detect LBW signals in experiments by irradiating a foil target with a single 10 PW-class laser. The distinct production locations of positrons in the LBW and BH processes lead to dramatically different acceleration dynamics and, consequently, to distinguishable positron spectra. The invariance of the high-energy tail of positron spectra with respect to the target thickness serves as a unique signature of LBW pair production, whereas BH positrons primarily occupy the intermediate-energy range and show a strong dependence on the target thickness. Additionally, in the presence of more intense lasers, LBW signals can also be separated from emerging NBW signals by introducing some preplasmas at the back of the target.

\begin{acknowledgments}

This work was supported by the National Science Foundation of China (Grants No. 12405285, 12135009 and 11991074) and the China Postdoctoral Science Foundation (Grants No.~2023M742294). The simulations were performed on the $\pi$ 2.0 supercomputer at Shanghai Jiao Tong University.

\end{acknowledgments}

\bibliography{reference}

\end{document}